\documentclass[reprint,
aps,pra,longbibliography,
10pt,
superscriptaddress,
amsmath,amssymb,
floatfix
]{revtex4-1} 

\usepackage{amsmath,amssymb,bm}
\usepackage{mathtools}
\usepackage{esvect}
\usepackage{booktabs}
\usepackage{tabularx}
\usepackage{array}
\usepackage{float}
\usepackage[svgnames]{xcolor}
\usepackage[colorlinks=true,
linkcolor=MediumBlue,
urlcolor=MediumBlue,
citecolor=MediumBlue]{hyperref}

\begin{document}

\title{Precision Mass and Density Measurement of Individual Optically Levitated Microspheres}

\author{Charles P. Blakemore}
\email{cblakemo@stanford.edu}
\affiliation{Department of Physics, Stanford University, Stanford, California 94305, USA}

\author{Alexander D. Rider}
\affiliation{Department of Physics, Stanford University, Stanford, California 94305, USA}

\author{Sandip Roy}
\affiliation{Department of Physics, Stanford University, Stanford, California 94305, USA}

\author{Alexander Fieguth}
\affiliation{Department of Physics, Stanford University, Stanford, California 94305, USA}

\author{Akio Kawasaki}
\affiliation{Department of Physics, Stanford University, Stanford, California 94305, USA}
\affiliation{W.~W.~Hansen~Experimental~Physics~Laboratory,~Stanford~University,~Stanford,~California~94305,~USA\looseness=-1}

\author{\\ Nadav Priel}
\affiliation{Department of Physics, Stanford University, Stanford, California 94305, USA}

\author{Giorgio Gratta}
\affiliation{Department of Physics, Stanford University, Stanford, California 94305, USA}
\affiliation{W.~W.~Hansen~Experimental~Physics~Laboratory,~Stanford~University,~Stanford,~California~94305,~USA\looseness=-1}

\date{\today}
\begin{abstract}

We report an \emph{in situ} mass measurement of approximately-$4.7{\text -}\mu$m-diameter, optically levitated microspheres with an electrostatic co-levitation technique. The mass of a trapped, charged microsphere is measured by holding its axial (vertical) position fixed with an optical feedback force, under the influence of a known electrostatic force. A mass measurement with $1.8\%$ systematic uncertainty is obtained by extrapolating to the electrostatic force required to support the microsphere against gravity in the absence of optical power. In three cases, the microspheres are recovered from the trap on a polymer-coated silicon beam and imaged with an electron microscope to measure their radii. The simultaneous precision characterization of the mass and radius of individual microspheres implies a density of $1.55\pm0.08~$g/cm$^3$. The ability to recover individual microspheres from an optical trap opens the door to further diagnostics. \\

\noindent DOI: \href{https://doi.org/10.1103/PhysRevApplied.12.024037}{10.1103/PhysRevApplied.12.024037}

\end{abstract}

\maketitle

\section{INTRODUCTION}
Optical trapping and manipulation of micron-sized dielectric particles in vacuum has been applied to optomechanics~\cite{Ashkin:1970,Ashkin:1971,Chang:2010,Li:2011,Hoang:2016,Monteiro:2018} and cavity cooling~\cite{Asenbaum:2013,Millen:2015,Fonseca:2016}, fundamental forces and interactions~\cite{Arvanitaki:2013,Moore:2014,Ether:2015,Rider:2016,Ranjit:2016,Jain:2016,Hempston:2017}, quantum mechanics~\cite{Romero:2010,Romero:2011}, quantum information~\cite{Gieseler:2012}, and surface science~\cite{Blakemore:2019}. In many of these applications, knowing the size, mass, and other characteristics of the trapped particles is critical to drawing conclusions about moments of inertia, optical spring constants, and force sensitivity. 

We present a technique to measure the mass of individually trapped microspheres (MSs), by balancing a known electrostatic force, the optical levitation force, and Earth's gravity. The electrostatic force is extrapolated to the condition of no optical power to determine the gravitational force on the MS, and thus infer its mass. This measurement requires fewer assumptions than other techniques~\cite{Chaste:2012,Ricci:2018} and is found to be independent of environmental conditions. The method is applicable to particles of any size, in any scattering regime, provided that a component of the optical power opposes gravity, and the direction of the gravitational field can be controlled.

Similar electrodynamic balances have been used to stably trap and levitate aerosol particles~\cite{Gobel:1997,Heinisch:2009,Hargreaves:2010,Krieger:2012,Singh:2017} as a platform for studying such things as droplet evaporation. It is possible to estimate charge-to-mass ratios for micron-sized aerosol particles thus levitated, but practical constraints severely limit both the precision and accuracy of these estimations, as discussed in Refs.~\cite{Krieger:2012,Singh:2017}.

It may be possible to derive a direct relation between the optical power required to levitate a MS at the center of the trap and the mass of the MS using numerical methods to develop solutions to the wave equations of Mie scattering theory~\cite{ott}. However, this requires a detailed understanding of the MS radius, nonsphericity, and index of refraction, as well as a full description of the optical potential in three dimensions. The technique described here bypasses these complications and their associated systematics, resulting in increased accuracy.

A subset of MSs are also individually collected from the optical trap with use of a mechanical probe and imaged via scanning electron microscopy (SEM) to determine their radii. Knowing both the mass and the radius of individual MSs, their density can be calculated. The radii determined from SEM images of those specific MSs are compared with the radii determined from SEM images of large populations of approximately 1000 MSs that have never been in the optical trap.

\section{EXPERIMENTAL SETUP}
The optical trap used here is described in Refs.~\cite{Rider:2018,Blakemore:2019}. Silica MSs obtained from the St\"ober process~\cite{Stober:1968,bangs_laboratories} with diameter of approximately $4.7{\text -}\mu$m are loaded into the trap by ejection from a vibrating glass slide placed above it. To efficiently load MSs, 1~mbar of residual gas is used to provide viscous damping. The chamber can then be evacuated to a final pressure of $10^{-6}~$mbar to reduce  thermal noise. Below 0.1~mbar, the trap requires active feedback for stabilization. The feedback system~\cite{Rider:2018} serves to provide viscous damping in all three degrees of freedom.  Importantly, the axial degree of freedom of the MS, stabilized by modulating the power of the trapping beam with an acousto-optic modulator, is held at a fixed position by proportional and integral feedback terms.

The loading procedure triboelectrically charges the MSs. A xenon flash-lamp, emitting ultraviolet light, is used to alter the MS charge state, $q_{\rm MS}$, over a wide range $-500e < q_{\rm MS} < 500e$, where $e$ is the elementary charge~\cite{Moore:2014,Rider:2016,Rider:2018,Blakemore:2019}. The arbitrarily set MS charge state, stable over timescales on the order of 1~month, is known with sub-$e$ precision, as individual quanta are added to or removed from a state of overall neutrality.

Charged MSs are shielded from external electric fields by a Faraday cage made of six electrodes, each with an independent bias voltage. The two electrodes directly above and below the trapped MS are used to generate a uniform, slowly varying electric field at the trap location, exerting an axial force on a charged MS. The relation between the applied voltage and the electric field within the trapping region is modeled by finite-element analysis, with an uncertainty that is much smaller than any other systematic uncertainty.

After measuring their mass, three MSs are collected on the end of a polymer-coated silicon beam, described in Refs.~\cite{Wang:2017,Blakemore:2019}, where they remain attached via van der Waals forces. Individual MSs are addressed to particular locations, recognizable from features on the silicon beam. The silicon beam is then removed from the chamber, and the three MSs are imaged by SEM to determine their individual radii. A population of MSs of the same variety and lot as those used in the trap are also measured by SEM. For this purpose, a monolayer of MSs is spread onto a silicon wafer and subsequently imaged by SEM. Various diffraction gratings~\cite{diffraction_gratings} are used to calibrate the instrument at each of the magnifications.

\section{MEASUREMENTS AND RESULTS}
\begin{figure}[t!]
\includegraphics[width=0.95\columnwidth]{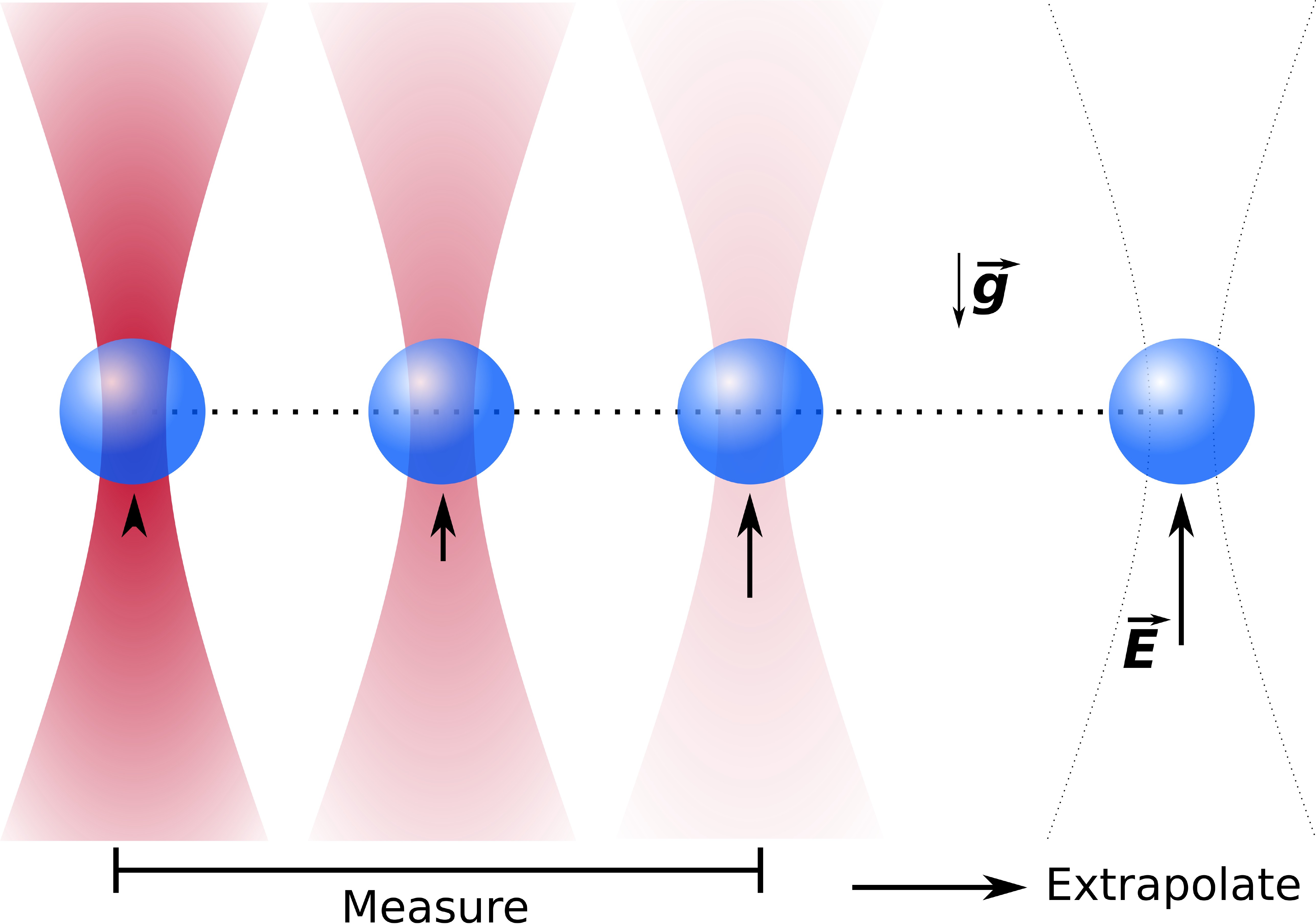}
\caption{\small Illustration of the measurement technique. A charged MS is trapped by a Gaussian laser beam and held at fixed axial position with active feedback. A slowly varying electric field is applied, depicted with a black arrow. The active feedback reduces the optical power, indicated by the intensity of the trapping beam, such that the sum of the optical and electrostatic levitation forces opposing gravity is constant. The relation between optical power and applied field is then extrapolated to zero optical power, allowing a determination of mass from the implied electrostatic levitation field and the known charge.}
\label{fig:schematic}
\end{figure}

Once a constant, known charge is obtained for a trapped MS, its axial position is fixed near the focus of the optical trap using the feedback. The slowly varying ($0.5~$Hz) electric field is applied in the vertical direction, while the power of the trapping beam injected into the chamber, controlled by the feedback, is monitored with a beam pickoff and a photodiode. As the applied electrostatic force increases, the axial feedback reduces the optical power required to maintain a net force of zero, counteracting gravity. The electric field can then be extrapolated to zero optical power, which allows a determination of the MS mass. This process is shown schematically in Fig.~\ref{fig:schematic}. The case of zero optical power cannot be directly measured, as there is a minimum power necessary both to constrain the MS to the optical axis, and to generate sufficient back-reflected light to measure the axial position via the methods described in Refs.~\cite{Rider:2018,Blakemore:2019}. The technique described is applicable only to single-beam traps~\cite{Rider:2018}, as its extension to systems with more than one beam requires care to account for the contributions of auxiliary beams to the total optical levitation force.

\begin{figure}[t!]
\includegraphics[width=0.95\columnwidth]{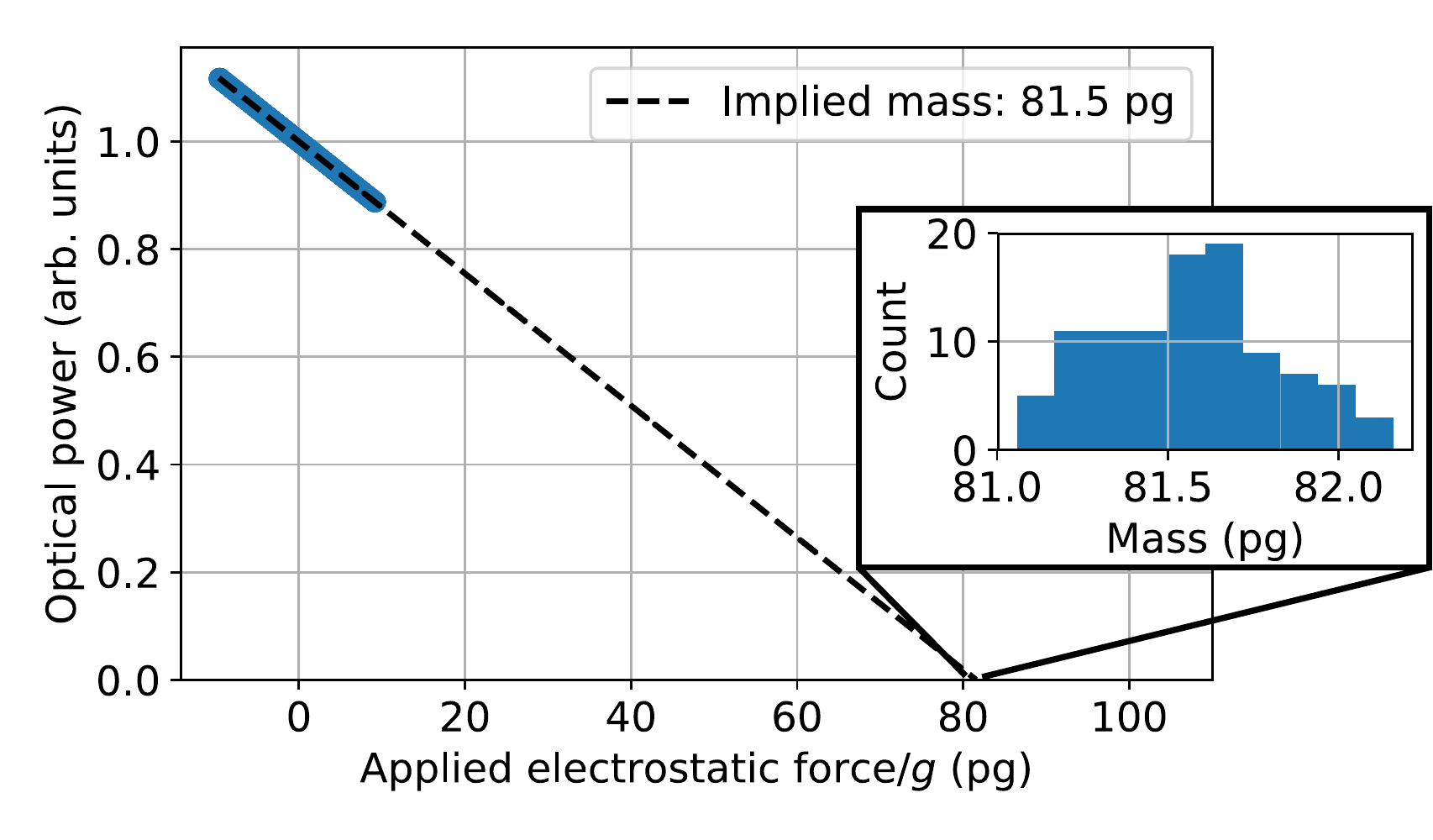}
\caption{ \small Normalized optical power versus applied electric field for 100 50-s integrations with a single MS. The extrapolation is performed separately for each integration. The mean of all extrapolations is shown with a dashed black line. The inset shows the distribution of the 100 extrapolated masses.}
\label{fig:example_data}
\end{figure}

The equilibrium of axial forces $F_z$ is expressed as,

\vspace*{-0.2cm}{}
\begin{equation} \label{eq:force_sum}
\begin{split}
\sum F_z &= q E(t) - m g + F_{{\rm opt},z} (t) = 0,
\end{split}
\end{equation}

\noindent where $q$ and $m$ are the charge and mass of the MS, respectively, $g=9.806~$m/s$^2$ is the local gravitational field strength~\cite{egm2008}, $F_{{\rm opt},z}(t)$ is the optical levitation force, assumed to be proportional to the trapping beam power, and $E(t)$ is the applied electric field strength. For each MS and charge-state combination, the slowly varying electric field and power are measured at least 50 times, each with a 50-s integration. An exemplary dataset is shown in Fig.~\ref{fig:example_data}, with the calculated masses from the extrapolation to zero optical power.

The extrapolation of this linear regression is performed over less than 1 order of magnitude and relies on only a few simple assumptions: firstly, the superposition principle, whereby the total force on the MS is the sum of gravity, the optical levitation force, and the electrostatic force, all of which are applied independently to the MS; secondly, proportionality between radiation pressure (the optical force) and photon flux, and thus the optical power~\cite{Jackson}; and finally, the linearity of the photodetection system, which we operate at a factor of 1000 below the manufacturer's reported saturation level~\cite{thorlabs}.

\begin{figure}[t!]
\includegraphics[width=0.95\columnwidth]{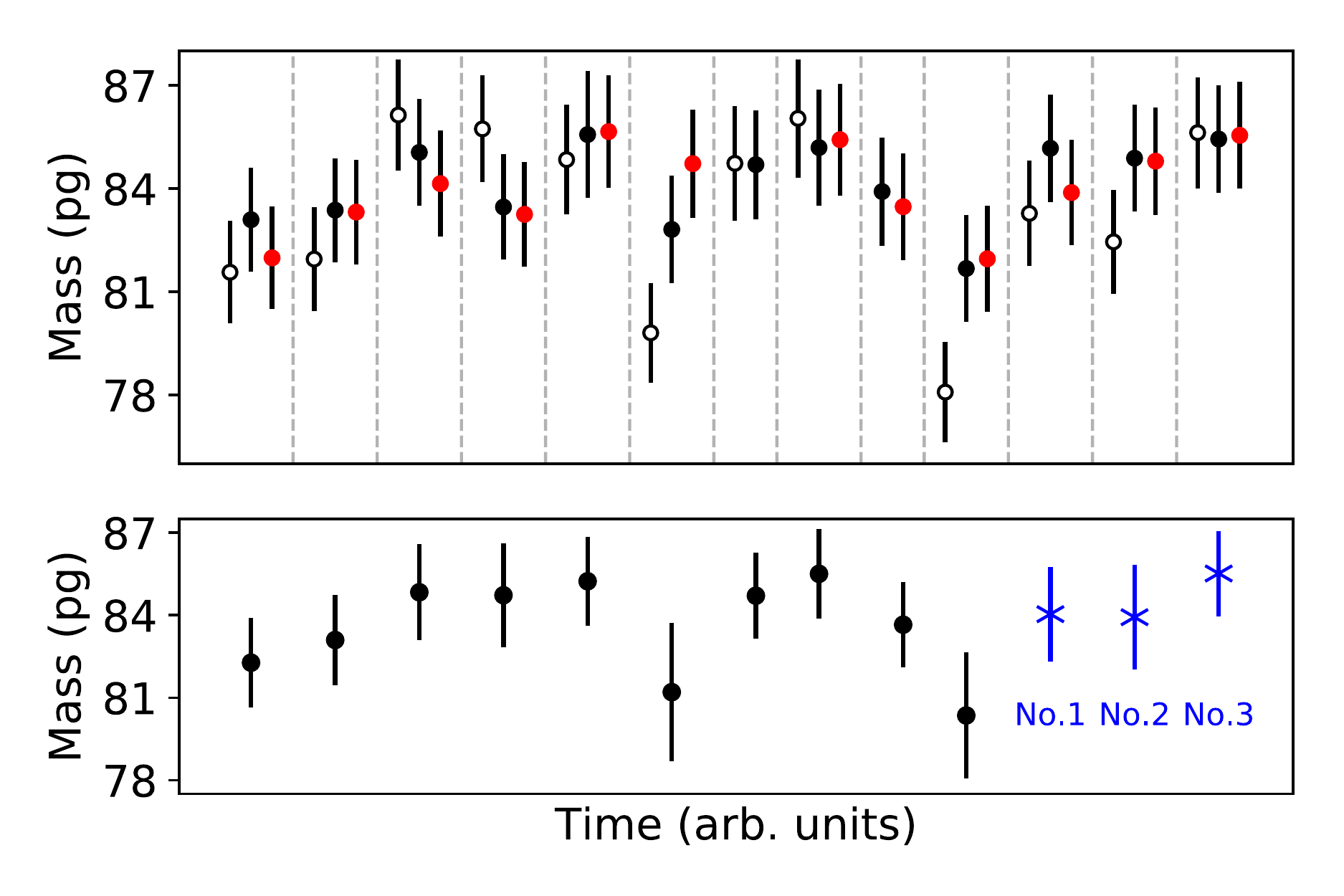}
\caption{ \small Measured MS masses in chronological order are shown in the top panel. Unfilled markers indicate a low-vacuum environment, $P=1.5~$mbar, while filled markers indicate high-vacuum environment, $P=10^{-6}-10^{-5}~$mbar. Black markers correspond to measurements with a negatively charged MS, while red markers correspond to measurements with a positively charged MS. Different MSs are separated by vertical dashed lines. The mean mass for each MS weighted over all experimental conditions is shown in the bottom panel. The blue data points with cross-shaped markers indicate the three MSs imaged by SEM following their mass measurement.}
\label{fig:mass_vs_time}
\end{figure}

The mass measurement is performed on 13 MSs in various charge states around $|q|=20e$, with both signs of charge, as well as in two vacuum-pressure regimes: trapping pressure, approximately $1~$mbar, and chamber base pressure, $10^{-5}~$mbar or less. The use of different pressures tests whether MS mass is lost due to heating, as reported for larger MSs in Ref~\cite{Monteiro:2017}. Cooling via residual gas decreases significantly with decreasing pressure, while absorption and scattering of laser light, the dominant heating mechanisms, remain constant. The results of mass measurements for all experimental conditions are shown in Fig.~\ref{fig:mass_vs_time}, while the results from the final three MSs later imaged by SEM are provided in the second column in Table~\ref{table:masses}.

To collect the final three MSs, the polymer-coated silicon beam is rapidly inserted between the trapping laser and the MS, allowing the MS to fall under the influence of gravity. Each distinct MS can be associated with the respective mass measurement given its position relative to the internal structure of the silicon beam. Van der Waals forces, enhanced by the polymer, serve to keep the MSs attached, whereas doped silicon and gold were both found to produce insufficient adhesion during previous attempts. The fluorocarbon polymer coating is made with a plasma-deposition technique inherent to the Bosch process~\cite{bosch}, with use of C$_4$F$_8$ and SF$_6$ gases in a $1.5$-kW inductively coupled plasma.

\begin{table}[b]\setlength{\hfuzz}{1.1\columnwidth}
\begin{minipage}{\textwidth}
  \caption{ \small MS masses, $m$, averaged over all experimental conditions; radii, $r$, averaged from two distinct high magnifications; and the derived density, $\rho$, for the three MSs caught on the silicon beam. All measurements include statistical and systematic uncertainties, and the relative contributions are shown explicitly for the measured masses.}
  \footnotesize
  \label{table:masses}
  \vspace*{1em}
  \setlength{\tabcolsep}{0.4em}
  {\renewcommand{\arraystretch}{1.4}
  \begin{tabular}{lccc}
    \toprule
    \hline
    \hline
    \small{MS} \hspace*{1cm} & \hspace*{3cm} \small{$m$ (pg)} \hspace*{3cm} & \hspace*{1cm} \small{$r$ ($\mu$m)} \hspace*{1cm} & \hspace*{1cm} \small{$\rho$ (g/cm$^3$)} \hspace*{1cm} \\
    \midrule
    \hline
    1 & $84.0 \pm 0.8{\rm\,(statistical)} \pm 1.5 {\rm\,(systematic)} $ & $2.348\pm0.038$ & $1.550\pm0.080$ \\
    2 & $83.9 \pm 1.1{\rm\,(statistical)} \pm 1.5 {\rm\,(systematic)} $ & $2.345\pm0.037$ & $1.554\pm0.079$ \\
    3 & $85.5 \pm 0.2{\rm\,(statistical)} \pm 1.5 {\rm\,(systematic)} $ & $2.355\pm0.038$ & $1.562\pm0.081$ \\
    \hline
    \hline
    \bottomrule
  \end{tabular}}
\end{minipage}
\end{table}

\begin{figure}[t!]
\includegraphics[width=1.0\columnwidth]{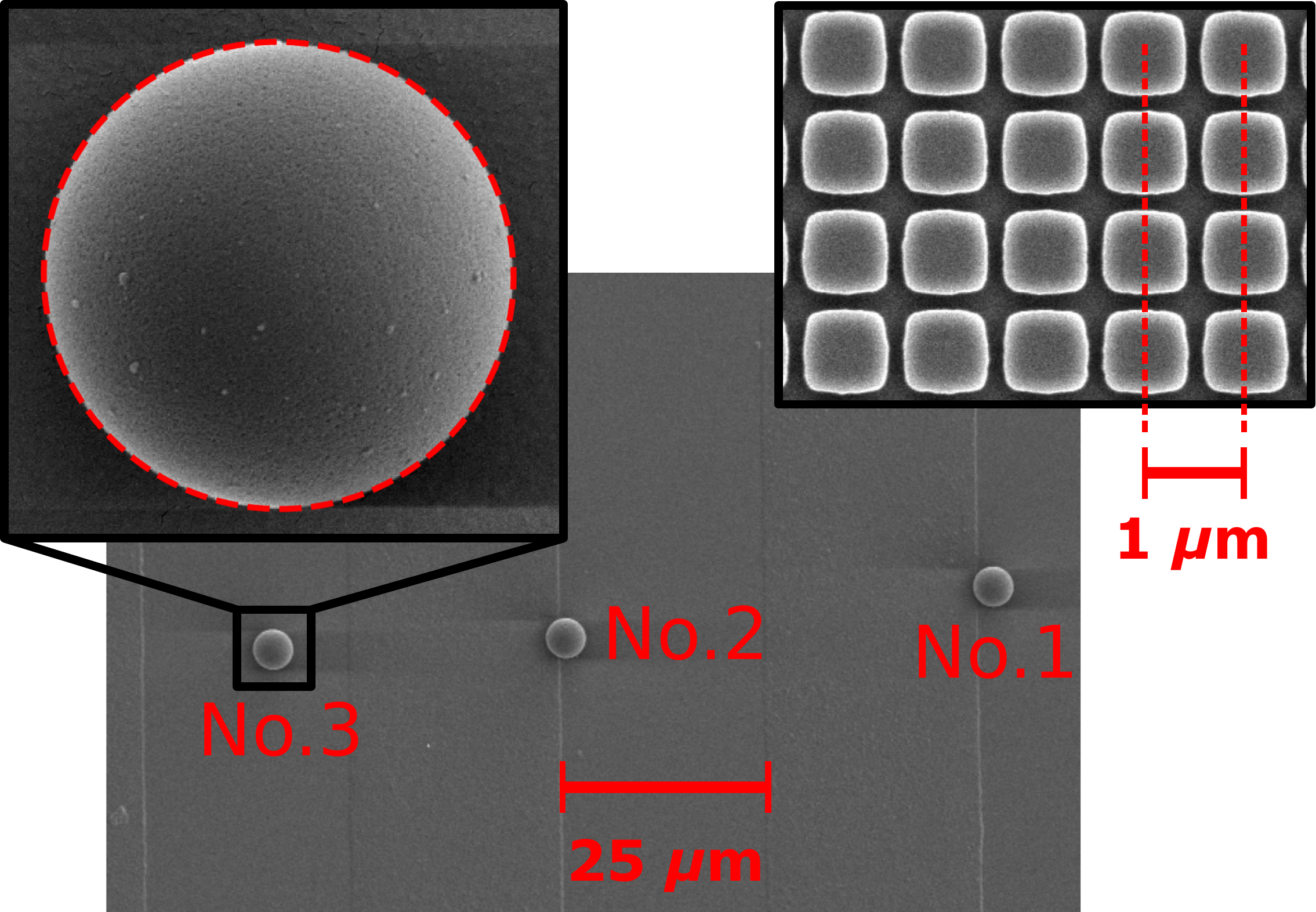}
\caption{ \small SEM images of the three MSs collected on the silicon beam, at ${\times}2500$ magnification. The left inset shows one MS at ${\times}35000$ magnification, overlaid with the best-fit ellipse, and the right inset shows the $1.000\pm0.005~\mu$m diffraction grating~\cite{diffraction_gratings}, also seen at ${\times}35000$ magnification. The diffraction grating serves as a calibration length scale for the high-magnification images of individual MSs.}
\label{fig:sem}
\end{figure}

For the SEM measurements, the silicon beam with three MSs is first sputter-coated with $100\pm50$~nm of a Au/Pd alloy to prevent charging and the resulting MS ejection from the silicon beam. Charging effects from the scanning electron microscope are significantly exacerbated by the nonconductive polymer, necessitating the relatively thick metal coating. A diffraction grating with a pitch of $1.000\pm0.005~\mu$m~\cite{diffraction_gratings} is used to calibrate the SEM images of individual MSs at high magnification, as seen in Fig.~\ref{fig:sem}.

The MS diameter is first determined in terms of raw pixels. This is done by edge detection and contour tracing to outline the MSs. The contour is then fit with an ellipse to account for real ellipticity in the MSs, as well as astigmatism in the electron microscope. The radius is taken as the average of the semimajor and semiminor axes, which differ by less than 1\%. A systematic uncertainty of $\pm1~$pixel in the determination of the semimajor and semiminor axes is included.

At the same level of magnification, images of the calibration grating are used for conversion from pixels to physical distances. This is done by locating the centroids of the grating's repeated structure in the image, and averaging the pixel distance between neighboring centroids across the entire image. The ratio of grating pitch in microns to observed grating pitch in pixels serves to calibrate the images. The $100\pm50$~nm thickness of the conductive coating is subtracted from the final radius.

To characterize the distribution of radii by SEM, a sparse monolayer of MSs is prepared on two heavily 

\newpage

\begin{figure}[b!]
\includegraphics[width=0.95\columnwidth]{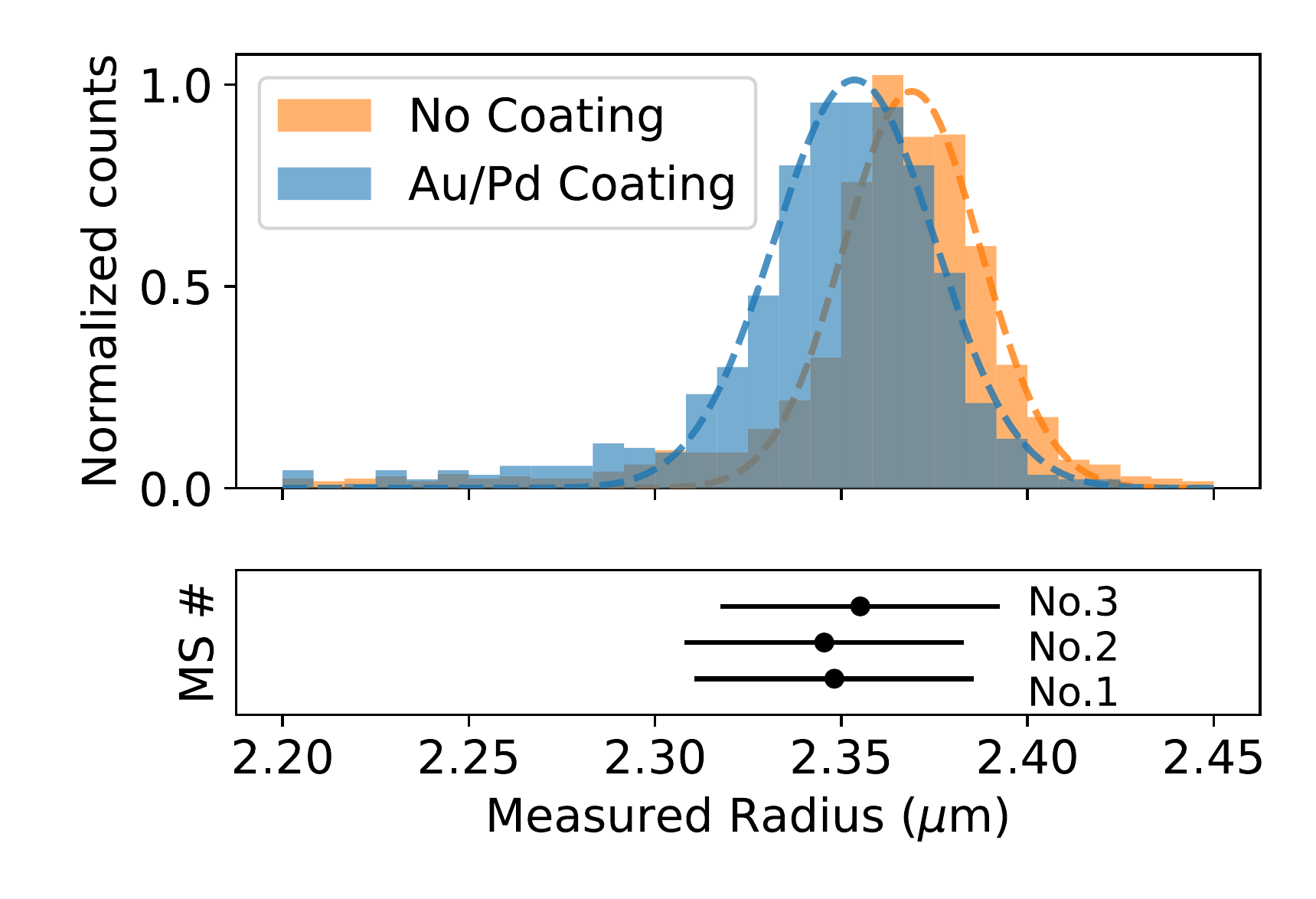}
\caption{ \small Distributions of MS radii measured by SEM for both conductively coated and uncoated MSs are shown in the top panel. Gaussian fits yield central values of $r_{\rm Au/Pd} = 2.35~\mu$m and $r_{\rm NC} = 2.37~\mu$m for the conductively coated and uncoated MSs, respectively. Each distribution is generated from approximately 1000 distinct MSs. Individual radius measurements from analysis of the MSs seen in Fig.~\ref{fig:sem} are shown in the bottom panel.}
\label{fig:radii}
\end{figure}

\noindent doped silicon wafers. Charging effects are reduced without the polymer, so a conductive coating is not strictly necessary. However, the two wafers are imaged with and without a conductive coating to study possible systematics. The coated wafer is sputtered with $40\pm10$~nm of the same Au/Pd alloy mentioned previously. Each wafer includes a diffraction grating with $9.98\pm0.02~\mu$m pitch~\cite{diffraction_gratings} to serve as a length calibration. Images of aapproximately 1000 distinct MSs, both conductively coated uncoated MSs, are collected at a range of magnifications, together with images of the calibration grating. The same ellipse identification and calibration procedures used for images of individual MSs are used to characterize the large MS populations.

The radii of the final three MSs measured are compared with the distribution of radii from the MS population measurements, shown in Fig.~\ref{fig:radii}. The conductive coating reduces the apparent size of the MSs by approximately $20$~nm after accounting for the coating thickness. This may be the result of charging of the uncoated MSs.

\section{DISCUSSION}

The technique described avoids a number of systematic uncertainties inherent to derivations of the MS mass from optical properties and the trapping potential~\cite{ott}. Importantly, the extrapolation to zero trapping-beam power is sensitive only to an offset in the power measurement, so an exact calibration of power is also unnecessary. Indeed, for the measurements reported here, the optical power is normalized to an arbitrary value of 1, as seen in Fig.~\ref{fig:example_data}. The only requirement for the measurement is that the photodiode responds linearly to the incident optical power, which is easily achieved with a device operating well below saturation.

Fluctuations in the mass measurement over the $50$ or more distinct $50$-s integrations for a set of experimental conditions are normally distributed with a standard deviation on the order of $0.5~$pg as seen in the inset in Fig.~\ref{fig:example_data}. However, the total uncertainty of the measurement is dominated by common systematic effects that are enumerated in Table~\ref{table:uncertainties}. Each effect listed is interpreted as an uncertainty on the applied electric field, the measured optical power, or the assumed value of $g$. 

From Eq.~(\ref{eq:force_sum}), these relative uncertainties directly propagate onto the extrapolated mass, whose uncertainty is computed as a quadrature sum of all contributions. The accuracy of the high-voltage amplifier's output monitor and the tolerance on the trapping lens focal length dominate the overall uncertainty. The second effect may offset the trap axially, thus sampling a different electric field strength. Each of the effects in Table~\ref{table:uncertainties} should result in a systematic shift common to all mass measurements. The total uncertainty obtained is $1.8\%$, which is included as a systematic uncertainty on the mean mass for each MS.

We also observe scatter in the measured mass of a single MS between different experimental conditions, as seen in Fig.~\ref{fig:mass_vs_time}. These variations could be due to a number of effects, including optical-path-length fluctuations in the axial feedback, electronic fluctuations in the axial feedback, and real changes in the mass of a MS. We do not observe any correlations between measured mass and experimental parameters such as the MS charge state or the vacuum pressure. The fluctuations are quantified by the standard deviation of measurements with different experimental conditions, which is included as part of the statistical uncertainty on the measured mass.

The measured masses and radii of the three imaged MSs are shown in Table~\ref{table:masses}, together with the calculated individual densities. This is consistent among the three MSs, and the average value, $\rho_{\rm MS} = 1.55 \pm 0.08~$g/cm$^3$, is significantly smaller than that of amorphous fused silica, $\rho_{\rm SiO_2} \approx 2.2~$g/cm$^3$~\cite{corning_silica}, as well as the value provided by the manufacturer, $\rho_{\rm Bangs} \approx 2.0~$g/cm$^3$~\cite{bangs_laboratories}. Other, indirect measurements of the density of silica nanoparticles~\cite{Ricci:2018} also differ significantly from the values reported here. This could be the result of nonidentical synthesis conditions and postsynthesis treatment by different manufacturers, which can have a large effect on final particle porosity~\cite{Li:2015,Kurdyukov:2018}.

\begin{table}[t!]
  \caption{ \small Systematic effects on the mass measurement. The amplifier discussed produces the voltage driving the electrodes, and thus the electric field. Geometric misalignment, including optical tolerances, can change the value of the electric field at the location of the trap. The application and subsequent measurement of the electric field is also subject to systematic effects, and each measurement channel can experience electrical pickup. Effects are determined empirically where possible, or are obtained from instrument specifications.}
  \label{table:uncertainties}
  \vspace*{1em}
  \setlength{\tabcolsep}{0.4em}
  {\renewcommand{\arraystretch}{1.1}
  \begin{tabular}{lcc}
    \toprule
    \hline
    \hline 
    \\[-8pt]
    Effect \, & \, & \, Uncertainty ($\times10^{-3}$) \\[2pt]
    \hline
    \midrule
    Amplifier-monitor accuracy$^{a}$ & & $\sigma_E / E \sim 15$ \\
    Lens focal length$^{a}$ & & $\sigma_E / E \sim 10$ \\
    Amplifier-gain uncertainty$^{a}$ & & $\sigma_E / E \sim 2$ \\
    Tilt of field axis & & $\sigma_E / E \sim 2$ \\
    Tilt of trap (optical) & & $\sigma_E / E \sim 1$ \\
    ADC offsets$^{a}$ & & $\sigma_{\mathcal{P}} / \mathcal{P} \sim 1$ \\
    Electrode-voltage offset & & $\sigma_E / E \sim 0.5$ \\
    dc-power offsets & & $\sigma_{\mathcal{P}} / \mathcal{P} \sim 0.3$ \\
    Local $g^{b}$ & & $\sigma_g / g \sim 0.1$ \\
    Fitting uncertainty & & $\sigma_{m} / m \sim 0.1$ \\
    Electrical pickup & & $\sigma_{\mathcal{P}} / \mathcal{P} \sim 0.02$ \\
    \hline
    \hline
    \bottomrule
  \end{tabular}}
  \begin{flushleft}
  ADC, analog-to-digital converter. \\
  $^{a}$ From manufacturer datasheets. \\
  $^{b}$ Estimated from Ref.~\cite{egm2008}.
  \end{flushleft}
\end{table}

Spin-echo small-angle neutron-scattering measurements on particles synthesized via the St\"ober have found an open-pore volume fraction of 32\% and an inaccessible-pore volume fraction of 10\% for particles with radius of approximately $80~$nm~\cite{Parnell:2016}. It is distinctly possible that MSs in solution absorb a nontrivial amount of water or other solvent, and that under low- to high-vacuum conditions, the liquid is removed, effectively lowering the mass and density. The classical electron oscillator model~\cite{Jackson} implies that the reduced density should result in a reduced refractive index: $n_{\rm MS}^2 - 1 = (n_{\rm SiO_2}^2 - 1) (\rho_{\rm MS} / \rho_{\rm SiO_2})$, leading to $n_{\rm MS} \approx 1.33$ at $1064~$nm, where $n_{\rm SiO_2}$ is the refractive index of fused silica~\cite{Malitson:1965}.

\section{CONCLUSION}
We present a technique using an electrodynamic balance together with an optical tweezer to precisely measure the gravitational mass of optically levitated silica microspheres. The measurement is limited by systematic uncertainties of approximately $1.8\%$ and is demonstrated to be independent of the (known) microsphere charge state, as well as the pressure of residual gas surrouding the microsphere.

After measuring their mass, three microspheres are collected from the trap with use of a mechanical probe and transferred to a scanning electron microscope, where their radii can be characterized. Together, the two precision characterizations allow direct calculation of the microsphere density. Large populations of microspheres used for trapping are also imaged. After accounting for the thickness of the coating, the individually measured radii of conductively coated microspheres are found to be consistent with the distribution of radii measured from the large population of conductively coated microspheres. 

The apparent independence of the measured mass on the vacuum pressure, as well as the consistency between the measured radii of individual microspheres that were optically trapped and large populations of micropsheres that were never trapped, both indicate a negligible loss of microsphere material by heating, under the environmental conditions tested for silica microspheres of radius approximately $2.35~\mu$m. The simplicity and accuracy of the mass measurement, along with the reliable transfer of specific microspheres from the optical trap to air and subsequently to a different vacuum environment, opens the possibility for other correlated, precision measurements on microscopic objects.

\section*{ACKNOWLEDGMENTS}
We would like to thank J.~Fox (Stanford University) for discussions on the readout electronics, R.G.~DeVoe (Stanford University) for a careful reading of the manuscript, and the Moore group (Yale University) for general discussions related to trapping microspheres. This work was supported, in part, by National Science Foundation Grants No.~PHY1502156 and No.~PHY1802952, Office of Naval Research Grant No.~N00014-18-1-2409, and the Heising-Simons Foundation.  A.K. acknowledges partial support from a William~M. and Jane~D. Fairbank Postdoctoral Fellowship of Stanford University. N.P. acknowledges the partial support from the Koret Foundation. Microsphere characterization was performed at the Stanford Nano Shared Facilities (SNSF), while silicon-beam fabrication was performed in part in the nano@Stanford laboratories, both of which are supported by the National Science Foundation as part of the National Nanotechnology Coordinated Infrastructure under award No.~ECCS-1542152.

\bibliography{mass_measurement}

\end{document}